\documentclass[showpacs,amsmath,amssymb,twoside,nofootinbib]{revtex4}
\usepackage[latin1]{inputenc}
\usepackage{bbm,mathrsfs}
\usepackage{hyperref}

\newcommand{\cc}{\text{\scshape c}}
\newcommand{\sm}{\text{\scshape sm}}

\begin{document}

\title{Sterile Neutrinos and Global Symmetries}

\author{J. Sayre, S.~Wiesenfeldt and S. Willenbrock}

\affiliation{Department of Physics, University of Illinois at
  Urbana-Champaign
  \\
  1110 West Green Street, Urbana, IL 61801, USA }

\begin{abstract}
  We use an effective-field-theory approach to construct models with
  naturally light sterile neutrinos, due to either exact or accidental
  global symmetries.  The most attractive models we find are based on
  gauge symmetries, either discrete or continuous.  We give examples
  of simple models based on $\mathbbm{Z}_\mathsf{N}$,
  $\mathsf{U(1)}^\prime$, and $\mathsf{SU(2)}^\prime$.
\end{abstract}


\maketitle


\section{Introduction}

A wide variety of neutrino-oscillation experiments indicate that
neutrinos have non-vanishing masses in the sub-eV range.\footnote{For
  a recent analysis, see Ref.~\cite{Strumia:2005tc}.}  All experiments
but one are consistent with three species of neutrinos: an atmospheric
pair with \mbox{$|\Delta m_{23}^2| \simeq 2.5\times
  10^{-3}\;\text{eV}^2$}, \mbox{$\sin 2\theta_{23} \simeq 1.0$}, and a
solar pair with \mbox{$|\Delta m_{12}^2| \simeq 8.0\times
  10^{-5}\;\text{eV}^2$}, \mbox{$\tan^2\theta_{12} \simeq 0.45$}. The
exception is the Liquid Scintillator Neutrino Detector (LSND)
experiment \cite{Aguilar:2001ty}, which finds \mbox{$|\Delta m^2|
  \simeq (0.2-7)\;\text{eV}^2$}, \mbox{$\sin2 \theta \simeq
  0.003-0.04$}.  This result requires one or more additional
neutrinos.  Since the decay $Z\to\nu\bar\nu$ shows that there are only
three species of neutrinos with ordinary weak interactions, these
additional neutrinos must be sterile with respect to the
standard-model gauge interactions.

In the standard model (SM), there are three species of active
neutrinos, and they are massless due to an accidental
$\mathsf{U(1)_{L}}$ global symmetry.\footnote{Throughout this paper,
  it is understood that all global symmetries are classical, and may
  have quantum anomalies.  These anomalies figure into the discussion
  in various places.  In the context of the SM, both $\mathsf{U(1)_L}$
  and $\mathsf{U(1)_B}$ are accidental global symmetries at the
  classical level, but only the combination $\mathsf{U(1)_{B-L}}$ is
  free of global anomalies.  Thus the neutrinos are exactly massless
  in the SM due to the accidental $\mathsf{U(1)_{B-L}}$ symmetry.}  If
one postulates, however, that there is new physics at a scale $M$ much
greater than the electroweak scale \mbox{$v \simeq 246\;\text{GeV}$},
it is natural for neutrinos to have small masses, independent of the
details of the new physics.  As long as the physics at $M$ does not
respect the accidental $\mathsf{ U(1)_{L}}$ symmetry of the SM, it may
give rise to a dimension-five interaction in the effective theory at
the weak scale proportional to $\frac{1}{M} \left( L\phi \right)
\left( L\phi \right)$, where $L$ and $\phi$ are the lepton and Higgs
doublets, respectively \cite{Weinberg:1979sa}. When the Higgs field
acquires a vacuum expectation value (vev) $v$, this interaction yields
neutrino masses of order $\frac{v^2}{M}$. For $M\sim
10^{14}-10^{16}\;\text{GeV}$, the resulting neutrino masses are in the
desired range.

The presence of the scale $M$ raises the question of why the other
fermions of the standard model do not acquire masses of order $M$.
The reason is that they are forbidden by the SM gauge interactions
from acquiring masses until the electroweak symmetry is broken at
the scale $v$.  Thus the other fermions of the SM have masses of
order $v$, regardless of the presence of new physics at the scale
$M$.

If sterile neutrinos exist in nature, they are not protected by SM
gauge interactions from acquiring masses of order $M$.  Thus, within
this framework, light ($m\sim\text{eV}$) sterile neutrinos are
unnatural.  For this reason there is considerable skepticism regarding
their existence. In addition, big bang nucleosynthesis favors just
three species of light neutrinos, although more than three is not
ruled out ($2.67 \le N_\nu \le 3.85$ at 68\% CL) \cite{Cyburt:2004yc}.
A global fit to the world's neutrino data favors at least two sterile
neutrinos \cite{Sorel:2003hf}. The MiniBooNE experiment at Fermilab
seeks to confirm the LSND result.

In this paper, we explore extensions of the SM in which sterile
neutrinos are light due to a global symmetry.  We use an
effective-field-theory approach, so our results are independent of the
details of the new physics that resides at the scale $M$.  We first
consider an exact global symmetry, and show that it is straightforward
to produce a model with light sterile neutrinos.  There is doubt,
however, that exact global symmetries exist in nature.  An exception
is a discrete symmetry that is the remnant of a broken gauge symmetry.
We then turn to approximate (accidental) symmetries, and show that
light sterile neutrinos require new gauge interactions and a set of
new fermions that are SM singlets. Thus an entire sterile sector is
required to explain the occurrence of light sterile neutrinos.

\section{Exact Global Symmetry \label{s:global}}

\subsection{Continuous Symmetry}

There is considerable doubt that exact global symmetries can exist in
nature \cite{Witten:2000dt}.  It is likely that such symmetries are
violated by quantum-gravitational processes \cite{Kallosh:1995hi}.
Furthermore, there are none known in nature. The only candidates are
$\mathsf{U(1)_{L}}$ and $\mathsf{U(1)_{B}}$, but these are accidental
global symmetries and are likely violated by higher-dimensional
operators. Indeed, this is the most compelling explanation of why
neutrinos are so much lighter than the other known fermions, as
discussed in the Introduction.

Despite these arguments, it remains a logical possibility that there
are exact global symmetries in nature that we have not yet discovered.
If a sterile neutrino $\nu_s$ is charged under this symmetry, it could
forbid a bare Majorana mass $m\nu_s\nu_s$. Here we consider the
possibility of an exact global $\mathsf{U(1)_X}$ symmetry, along with
the SM gauge group \mbox{$\mathsf{G_\sm=SU(3)_C \times SU(2)_L \times
    U(1)_Y}$}, and assign a charge of unity to the sterile neutrino,
$\mathsf{X}_{\nu_s}=1$.

In order to explain why the sterile neutrino is much lighter than the
weak scale, we need the $\mathsf{U(1)_X}$ symmetry to forbid the
operator $\left( L\phi \right) \nu_s$.\footnote{All operators
  considered in this paper are constructed from a pair of left-handed
  Weyl fermions combined to form a Lorentz scalar, plus additional
  scalar fields.  The $\mathsf{SU(2)_L}$ fields contained in
  parentheses form an $\mathsf{SU(2)_L}$ singlet.}  Such an operator
would generate a Dirac neutrino mass at the weak scale.  The symmetry
$\mathsf{U(1)_{L}}$ allows this operator (with
$\mathsf{L}_{\nu_s}=-1$), so it cannot play the role of
$\mathsf{U(1)_X}$.  We want the sterile neutrino to acquire mass via a
dimension-five operator, analogous to the operator \mbox{$\left( L\phi
  \right) \left( L\phi \right)$} that generates a small mass for the
active neutrinos.  The only candidate operators are
$\nu_s\nu_s\left(\phi\phi\right)$ and $\nu_s\nu_s(\phi\tilde\phi)$,
where $\tilde\phi\equiv\epsilon\phi^*$; the former is forbidden by
$\mathsf{U(1)_Y}$, the latter by $\mathsf{U(1)_X}$ regardless of the
charge carried by $\phi$.

Thus, in order to proceed on this tack, we must extend the Higgs
sector.  Adding a second Higgs doublet $\phi_2$ with hypercharge
$-\frac12$ would allow the operator
$\nu_s\nu_s\left(\phi_1\phi_2\right)$, which would generate a small
Majorana mass for $\nu_s$ if both $\phi_1$ and $\phi_2$ acquire
weak-scale vevs.  However, we also need order-unity mixing between
active and sterile neutrinos, which requires a dimension-five operator
of the form $L\nu_s\times\left(\text{two Higgs fields}\right)$.  Such
an operator is forbidden by the $\mathsf{SU(2)_L}$ gauge symmetry if
the only Higgs fields available are $\mathsf{SU(2)_L}$ doublets.

Rather than adding a second Higgs doublet, let's add a Higgs singlet
$S$ -- a sterile Higgs. The effective Lagrangian is\footnote{This
  Lagrangian may be generalized by adding additional powers of $S/M$
  to each of the operators \cite{Langacker:1998ut}. The resulting
  neutrino masses would then be much less than the eV scale.}
\begin{align} \label{eq:l-global}
  \mathscr{L} & = \frac{c_1}{M} \left( L\phi \right) \left( L\phi
  \right) + \frac{c_2}{M}\, \nu_s\nu_s S S + \frac{c_3}{M} \left(
    L\phi \right) \nu_s S + \text{h.c.} \ ,
\end{align}
where $c_j$ are dimensionless coefficients. The second term requires
that the Higgs field carries $\mathsf{U(1)_X}$ charge
$\mathsf{X}_S=-1$. The other terms require that $L$ and $\phi$ carry
equal and opposite $\mathsf{U(1)_X}$ charge. Also taking into account
the Yukawa couplings of the quarks and leptons, we list in
Table~\ref{tb:u1charges} the $\mathsf{U(1)_X}$ charges of the fields
of the SM as well as of the sterile neutrino and sterile Higgs.
%
\begin{table}[t]
  \centering
  \begin{tabular}{ll}
    field & $\mathsf{U(1)_X}$ charge \\
    \hline
    $Q$ & $\phantom{-}\mathsf{X}_Q$ \\
    $u^\cc$ & $-\mathsf{X}_Q + \mathsf{X}_L$ \\
    $d^\cc$ & $-\mathsf{X}_Q - \mathsf{X}_L$ \\
    $L$ & $\phantom{-}\mathsf{X}_L$ \\
    $e^\cc$ & $-2\,\mathsf{X}_L$ \\
    $\nu_s$ & $\phantom{-}1$ \\
    $\phi$ & $-\mathsf{X}_L$ \\
    $S$ & $-1$
  \end{tabular}
  \caption{A simple model for light sterile neutrinos based on an
    exact global $\mathsf{U(1)_X}$ symmetry. \label{tb:u1charges}}
\end{table}
%
There are two free parameters, namely the charges of the quark and
lepton doublets relative to the charge of the sterile neutrino
(which we have normalized to unity).

When the sterile Higgs acquires a vev, it breaks the $\mathsf{U(1)_X}$
symmetry and generates a Majorana mass for the sterile neutrino via
the second term in Eq.~(\ref{eq:l-global}).  Since it is a sterile
Higgs field, it does not break the electroweak symmetry.  Thus its vev
is not required to lie at the weak scale.  We only choose it to do so
in order to generate sterile neutrino masses in the desired range.

The breaking of the global $\mathsf{U(1)_X}$ symmetry produces a
Goldstone boson, corresponding to the phase of the sterile Higgs field
$S$.  This Goldstone boson couples to both active and sterile
neutrinos with strength $\frac{v}{M}$ via the last two interaction
terms of Eq.~(\ref{eq:l-global}).  Such a weak coupling means that the
Goldstone bosons are not in thermal equilibrium with the neutrinos
during the era of nucleosynthesis.\footnote{Goldstone bosons with
  stronger couplings can have observable consequences for cosmology
  \cite{Chacko:2003dt, Chacko:2004cz, Hall:2004yg, Beacom:2004yd}.}
We therefore cannot calculate their contribution to the energy density
of the universe, and hence to the expansion rate, during
nucleosynthesis.  This is in contrast to the sterile neutrinos
themselves, which are in thermal equilibrium via their mixing with
active neutrinos, and contribute to the expansion rate in a known way
\cite{Shi:1993hm, Dolgov:2002wy, Cirelli:2004cz}.

The coupling of active and sterile neutrinos to the Goldstone boson
produces neutrino decay with a lifetime of order \mbox{$\tau^{-1}\sim
  \frac{1}{16\pi}\;m_\nu\left(\frac{v}{M}\right)^2 \sim
  \left(10^{2-6}\;\text{yrs}\right)^{-1}$}.  This has potentially
observable consequences for neutrinos emitted from astrophysical
sources \cite{Beacom:2002vi}.

We have already mentioned that it is problematic to have an exact
global symmetry in the presence of gravity.  Even putting that aside,
the $\mathsf{U(1)_X}$ symmetry above suffers from global anomalies
that make it difficult to interpret it as a fundamental symmetry of
nature.  Consider, for example, the mixed
$\mathsf{U(1)_X}$-gravitational anomaly, proportional to
\begin{align} \label{eq:grav-anomaly}
  3\cdot 2\cdot \mathsf{X}_Q + 3 \left( - \mathsf{X}_Q + \mathsf{X}_L
  \right) + 3 \left( - \mathsf{X}_Q - \mathsf{X}_L \right) + 2\cdot
  \mathsf{X}_L - 2\, \mathsf{X}_L + 1 & = 1 \ .
\end{align}
Every term cancels except the one provided by the sterile
neutrino.

Let us consider the other global anomalies.  The $\left[
  \mathsf{SU(3)_C} \right]^2 \mathsf{U(1)_X}$ anomaly cancels for any
value of $\mathsf{X}_L$ and $\mathsf{X}_Q$.  The cancellation of the
$\left[ \mathsf{SU(2)_L} \right]^2 \mathsf{U(1)_X}$ anomaly requires
$\mathsf{ X}_L=-3\,\mathsf{X}_Q$.  The $\mathsf{U(1)_X}$ charges of
the SM particles are then just proportional to their hypercharges.
Thus the $\left[ \mathsf{U(1)_Y} \right]^2 \mathsf{U(1)_X}$ anomaly
cancels, since the sterile neutrino does not contribute to it.

Simply adding another sterile neutrino with the opposite
$\mathsf{U(1)_X}$ charge is enough to cancel all global anomalies;
however, this is not an acceptable solution, as there is no symmetry
preventing the two sterile neutrinos from pairing up to form a Dirac
neutrino of mass $M$.  The minimum number of sterile particles needed
to both cancel all global anomalies and to form a chiral
representation of $\mathsf{U(1)_X}$ is thus three, with charges
$1,\,\mathsf{X}_s,-1-\mathsf{X}_s$ ($\mathsf{X}_s \neq -1$).

\subsection{Discrete Symmetry \label{s:discrete}}

Thus far we have considered an exact continuous global symmetry to
explain why sterile neutrinos are light; however, an exact discrete
symmetry could also be employed for this purpose.  For example, any
$\mathbbm{Z}_\mathsf{N}$ subgroup ($\mathsf{N}>2$) of
$\mathsf{U(1)_X}$ would be sufficient to forbid a sterile neutrino
bare mass.  In this scenario $\nu_s$ carries charge 1 and $S$ carries
charge $-1+\mathsf{N}$, so $\nu_s S=0$ mod \textsf{N} and the Yukawa
interactions of Eq.~(\ref{eq:l-global}) are allowed. SM particles
could also carry nonzero $\mathbbm{Z}_\mathsf{N}$ charges.

As in the case of a continuous global symmetry, it is widely believed
that a discrete symmetry is violated by quantum-gravitational effects.
However, unlike the case of a continuous global symmetry, a discrete
symmetry can be protected from quantum-gravitational effects if it is
a subgroup of a broken gauge symmetry (``discrete gauge symmetry'')
\cite{Krauss:1988zc}.  For completeness, we will not restrict our
discussion to a discrete gauge symmetry until the end of this section.

The discrete symmetry is broken by the vev of $S$ without producing a
Goldstone boson; however, a broken discrete symmetry gives rise to
domain walls that are in conflict with cosmological observations
\cite{Preskill:1991kd}. There are at least two ways to avoid this
problem.  The first is for the discrete symmetry to be anomalous,
thereby lifting the degeneracy among the vacuum states
\cite{Preskill:1991kd}.  However, there are two potential pitfalls
with this solution.  The first is that if the discrete symmetry is
anomalous, then it cannot be a discrete gauge symmetry, and hence it
may be violated by quantum-gravitational effects.  The second is that,
as in the case of a continuous global symmetry, it is difficult to
interpret an anomalous discrete symmetry as a fundamental symmetry.

The second way to avoid problems with domain walls is for the discrete
symmetry to be a discrete gauge symmetry.  This solves the problem if
the gauge symmetry is broken after the era of inflation.  If it is
broken before the era of inflation, then there is no mechanism to
remove the domain walls \cite{Preskill:1991kd}.

Let us consider a simple anomaly-free discrete symmetry.  The simplest
candidate is $\mathbbm{Z}_3$, with $\nu_s$ carrying charge 1 and $S$
carrying charge 2.  However, in addition to the Yukawa interactions of
Eq.~(\ref{eq:l-global}), this symmetry allows the dimension-four
operator $\nu_s\nu_sS^*$, which would generate a weak-scale mass for
the sterile neutrino when $S$ acquires a vev. Thus this model is not
viable.

The next simplest candidate is $\mathbbm{Z}_4$, where $\nu_s$ carries
charge 1 and $S$ carries charge $3$.  In addition to the Yukawa
interactions of Eq.~(\ref{eq:l-global}), the dimension-five operator
\mbox{$\nu_s \nu_s S^* S^*$} is also allowed.  This operator also
contributes to an eV-scale mass for the sterile neutrino when $S$
acquires a vev.  Hence this is the simplest viable model.  A single
sterile neutrino has a discrete gauge anomaly, so one must add
additional sterile neutrinos.  The simplest anomaly-free model has two
sterile neutrinos with unit $\mathbbm{Z}_4$ charge.  This satisfies
the discrete gravitational anomaly, which requires that the sum of the
charges equal 0 mod \textsf{N} (mod \textsf{N}/2 for \textsf{N} even)
\cite{Ibanez:1991hv,Ibanez:1992ji}. The other anomaly conditions place
no constraint on this model.

The $\mathbbm{Z}_4$-symmetric sterile-Higgs potential is given by
\begin{align} \label{eq:l-potential}
  V & = -\mu^2 S^* S + \lambda_1 \left( S^* S \right)^2 + \lambda_2
  \left( S^4 + S^{*4} \right) \ .
\end{align}
When the sterile Higgs field acquires a vev, it breaks the
$\mathbbm{Z}_4$ symmetry and generates sterile neutrino masses and
mixing with active neutrinos via Eq.~(\ref{eq:l-global}) and the
additional $\mathbbm{Z}_4$-symmetric Yukawa interaction
$\nu_s\nu_sS^*S^*$ mentioned above.

There are many other possibilities for $\mathsf{N>4}$.  Another simple
anomaly-free model that yields two eV-scale sterile neutrinos is
$\mathbbm{Z}_6$ with fermions of charge $(1,2)$ and a sterile Higgs
field of charge $5$.  Both fermions acquire mass via dimension-five
operators, $\psi_1\psi_1S_5S_5$ and $\psi_2\psi_2S^*_1S^*_1$.  Only
$\psi_1$ mixes with the active neutrinos via a dimension-five
operator, $(L\phi)\psi_1S_5$.  We sketch a simple gauge theory that
gives rise to such a $\mathbbm{Z}_6$ model in an Appendix.

In this section we have explored the possibilities for explaining the
presence of light sterile neutrinos via exact global symmetries,
either continuous or discrete.  While it is easy to construct such
models, they can all be criticized in one way or another.  The only
model that has no shortcomings, discussed in the paragraphs above, was
one based on a discrete gauge symmetry.

\section{Accidental Global Symmetry \label{s:acc}}

\subsection{$\mathsf{U(1)}^\prime$ \label{s:u1}}

If $\mathsf{U(1)_X}$ were a local rather than a global symmetry, then
the absence of a bare Majorana mass term $m\nu_s\nu_s$ would be
ascribed to an accidental flavor symmetry, call it sterile neutrino
number, $\mathsf{U(1)_S}$, corresponding to $\nu_s\to
e^{i\theta}\nu_s$.  This symmetry is explicitly violated by the last
two dimension-five operators of Eq.~(\ref{eq:l-global}), and is
therefore only approximate.  This is completely analogous to the case
of the active neutrinos, with $\mathsf{U(1)_S}$ playing the role of
$\mathsf{U(1)_{L}}$.  Unfortunately, this rather elegant model has a
fatal flaw: the $\mathsf{U(1)_X}$ local symmetry is anomalous, as
discussed in the previous section.  If we choose
$\mathsf{X}_L=-3\,\mathsf{X}_Q$, then the $\mathsf{U(1)_X}$ charges of
all SM particles are proportional to their hypercharges, and most
anomalies cancel, as discussed above.  The only nonvanishing anomalies
are the mixed $\mathsf{U(1)_X}$-gravitational anomaly and the $\left[
  \mathsf{U(1)_X} \right]^3$ anomaly.  The contribution of SM
particles to these anomalies vanish, so anomaly cancellation must
occur in the sterile sector separately.\footnote{This conclusion
  depends on the requirement that neutrino masses and mixings arise
  from the dimension-five operators of Eq.~(\ref{eq:l-global}). If one
  relaxes that constraint, then one may construct
  $\mathsf{U(1)}^\prime$ models in which anomaly cancellation occurs
  between SM and sterile fermions \cite{Appelquist:2002mw}.}

As in the previous section, simply adding another sterile neutrino
with the opposite $\mathsf{U(1)_X}$ charge is enough to cancel all
anomalies; however, this is not an acceptable solution, as there is no
symmetry preventing the two sterile neutrinos from pairing up to form
a Dirac neutrino of mass $M$.  The minimum number of sterile particles
needed to both cancel all gauge anomalies and to form a chiral
representation of $\mathsf{U(1)_X}$ is five \cite{Babu:2003is}.  Thus
there must exist an entire sterile sector.  SM particles may carry
$\mathsf{U(1)_X}$ charges (proportional to their hypercharges), or
they may be $\mathsf{U(1)_X}$ singlets (corresponding to
$\mathsf{X}_L=\mathsf{X}_Q=0$).

The $\mathsf{U(1)_X}$ charges of the five sterile particles are not
restricted to commensurate values.  If we impose such a restriction,
the simplest model (in the sense of having the smallest integer
charges) is $\mathsf{U(1)_X}=\left(-8,-7,1,5,9\right)$
\cite{Davoudiasl:2005ks}. This is much simpler than the solution given
in Ref.~\cite{Babu:2003is}. We list in Table~\ref{tb:u1-charges} the
fifty simplest integer solutions; that of Ref.~\cite{Babu:2003is} is
number 42 in the list.  The charges must be commensurate if
$\mathsf{U(1)_X}$ is a subgroup of a simple gauge group
\cite{Georgi:1974sy}.

\begin{table}[t]
  \begin{tabular}{|r|rrrrr|p{6pt}|r|rrrrr|p{6pt}|r|rrrrr|}
    \cline{1-6}
    \cline{8-13}
    \cline{15-20}
    1 & -8 & -7 & 1 & 5 & 9 & &
    18 & -33 & -32 & 5 & 21 & 39 &  &
    35 & -50 & -27 & 11 & 14 & 52 \\
    2 & -9 & -7 & 2 & 4 & 10 &  &
    19 & -34 & -18 & 4 & 13 & 35 & &
    36 & -50 & -41 & 9 & 26 & 56 \\
    3 & -18 & -17 & 1 & 14 & 20 & &
    20 & -35 & -22 & 2 & 19 & 36 & &
    37 & -50 & -42 & 12 & 23 & 57 \\
    4 & -21 & -12 & 5 & 6 & 22 &  &
    21 & -39 & -30 & 4 & 23 & 42 & &
    38 & -50 & -49 & 1 & 44 & 54 \\
    5 & -22 & -18 & 7 & 8 & 25 &  &
    22 & -44 & -25 & 9 & 14 & 46 & &
    39 & -51 & -39 & 9 & 25 & 56 \\
    6 & -22 & -20 & 7 & 9 & 26 &  &
    23 & -45 & -32 & 11 & 17 & 49 & &
    40 & -51 & -40 & 13 & 21 & 57 \\
    7 & -25 & -14 & 4 & 9 & 26 & &
    24 & -45 & -39 & 3 & 32 & 49 & &
    41 & -52 & -50 & 17 & 22 & 63 \\
    8 & -25 & -17 & 7 & 8 & 27 & &
    25 & -45 & -42 & 6 & 29 & 52 & &
    42 & -54 & -23 & 8 & 14 & 55 \\
    9 & -25 & -23 & 2 & 18 & 28 & &
    26 & -46 & -35 & 12 & 18 & 51 & &
    43 & -55 & -23 & 9 & 13 & 56 \\
    10 & -26 & -14 & 5 & 8 & 27 & &
    27 & -46 & -45 & 13 & 22 & 56 & &
    44 & -55 & -36 & 7 & 26 & 58 \\
    11 & -26 & -18 & 5 & 11 & 28 & &
    28 & -48 & -30 & 13 & 14 & 51 & &
    45 & -55 & -49 & 13 & 27 & 64 \\
    12 & -27 & -25 & 7 & 13 & 32 & &
    29 & -48 & -42 & 9 & 26 & 55 & &
    46 & -56 & -25 & 6 & 18 & 57 \\
    13 & -28 & -26 & 6 & 15 & 33 & &
    30 & -48 & -42 & 6 & 30 & 54 & &
    47 & -56 & -34 & 5 & 27 & 58 \\
    14 & -28 & -27 & 10 & 11 & 34 & &
    31 & -49 & -31 & 10 & 18 & 52 & &
    48 & -57 & -45 & 17 & 21 & 64 \\
    15 & -32 & -23 & 9 & 11 & 35 & &
    32 & -49 & -36 & 1 & 34 & 50 & &
    49 & -63 & -25 & 11 & 13 & 64 \\
    16 & -32 & -31 & 1 & 27 & 35 & &
    33 & -49 & -41 & 8 & 27 & 55 & &
    50 & -63 & -29 & 5 & 23 & 64 \\
    17 & -33 & -24 & 3 & 19 & 35 & &
    34 & -49 & -47 & 2 & 40 & 54 & &
    & & & & & \\
    \cline{1-6}
    \cline{8-13}
    \cline{15-20}
  \end{tabular}
  \caption{$\mathsf{U(1)_X}$ charges of the additional five sterile
    particles that satisfy the anomaly conditions, $\sum
    \mathsf{X}_i=0$ and $\sum \mathsf{X}_i^3=0$, for integer values. 
    \label{tb:u1-charges}}
\end{table}

In the absence of Yukawa couplings, all five sterile particles are
massless, and carry an accidental $\left[\mathsf{U(1)}\right]^5$
global symmetry. In order to generate masses for the particles, we
must explicitly violate this global symmetry with Yukawa couplings, as
well as break the $\mathsf{U(1)_X}$ gauge symmetry. The pattern of
fermion masses depends on the $\mathsf{U(1)_X}$ charge of the Higgs
field or fields that are chosen.

Let us consider an explicit model as an illustration.  We choose the
simplest model, with $\mathsf{U(1)_X}$ charges
$\left(-8,-7,1,5,9\right)$.  In order to make $\nu_sS$ invariant under
$\mathsf{U(1)_X}$, we must choose the $\mathsf{U(1)_X}$ charge of $S$
to be opposite of one of the five fermions.  Let us consider the case
where the Higgs field carries charge $-1$. Including terms up to
dimension five, the Yukawa interactions allowed by the gauge symmetry
are
\begin{multline} \label{eq:l-five}
  \mathscr{L} = y\,\psi_{9} \psi_{-8} S_{-1} + \frac{c_1}{M} \left(
    L\phi \right) \left( L\phi \right) + \frac{c_2}{M}\, \psi_{1}
  \psi_{1} S_{-1} S_{-1} + \frac{c_3}{M} \left( L\phi \right) \psi_{1}
  S_{-1}
  \\
  + \frac{c_4}{M}\, \psi_{9} \psi_{-7} S_{-1} S_{-1} + \frac{c_5}{M}\,
  \psi_{5} \psi_{-7} S^*_{1} S^*_{1} + \text{h.c.} \ ,
\end{multline}
where the subscript on the fields indicate their $\mathsf{U(1)_X}$
charge.  The field $\psi_{1}$ plays the role of the sterile neutrino,
$\nu_s$, in Eq.~(\ref{eq:l-global}), and the field $S_{-1}$ plays the
role of $S$.  When the sterile Higgs field $S_{-1}$ acquires a
weak-scale vev, the fields $\psi_9, \psi_{-8}, \psi_5, \psi_{-7}$ form
Dirac neutrinos with masses of order $v$ and $\frac{v^2}{M}$. Neither
of these Dirac neutrinos couples to the active neutrinos.  Thus this
model yields one light sterile neutrino that mixes with the active
neutrinos.  The light Dirac neutrino does not contribute in a
calculable way to the expansion rate during big bang nucleosynthesis
because it is not in thermal equilibrium with the active neutrinos.

Any of the models listed in Table~\ref{tb:u1-charges} can be used to
generate light sterile neutrinos along the lines above, given a
suitable choice of sterile Higgs field(s).  However, none of these
models is particularly compelling, in that they are not motivated by
anything other than the desire to produce naturally light sterile
neutrinos.

\subsection{$\mathsf{SU(2)}^\prime$}

Let us consider replacing the $\mathsf{U(1)_X}$ symmetry with
$\mathsf{SU(2)}^\prime$.  The sterile neutrino must transform
nontrivially under this symmetry.  In order to avoid having to add yet
another particle to the model, let its $\mathsf{SU(2)}^\prime$ partner
be the positron field $e^\cc$.  This is the left-right symmetric model
with the gauge group \mbox{$\mathsf{SU(3)_C \times SU(2)_L \times
    SU(2)_R \times U(1)_{B-L}}$} \cite{Pati:1974yy}.\footnote{We do
  not impose any additional discrete symmetries.}  The particle
content of this anomaly-free model is given in
Table~\ref{tb:lr-model}.

\begin{table}[t]
  \centering
  \begin{tabular}{lccl}
    field & $\mathsf{SU(2)_L}$ & $\mathsf{SU(2)_R}$ &
    $\mathsf{U(1)_{B-L}}$
    \\
    \hline
    $Q$ & 2 & 1 & $\phantom{-}\frac13$ \\
    $Q^\cc$ & 1 & 2 & $-\frac13$ \\
    $L$ & 2 & 1 & $\phantom{-}1$ \\
    $L^\cc$ & 1 & 2 & $-1$ \\
  \end{tabular}
  \caption{Particle content and gauge charges of the left-right
    symmetric model. \label{tb:lr-model}}
\end{table}

In order to generate quark masses, we need a Higgs field $\Phi$ in the
$\left(2,2\right)$ representation of $\mathsf{SU(2)_L\times SU(2)_R}$,
which gives rise to the Yukawa coupling
\begin{align} \label{eq:l-leftright-quark}
  \mathscr{L} & = y_Q\, Q\, \Phi\, Q^\cc \ ,
\end{align}
where $Q^\cc$ is an $\mathsf{SU(2)_R}$ doublet consisting of
$\left(u^\cc,d^\cc\right)$.  This coupling generates masses for
up-type and down-type quarks when the diagonal components of $\Phi$
acquire vevs.  Unfortunately, the operator
\begin{align} \label{eq:l-leftright-lepton}
  \mathscr{L} & = y_L\, L\, \Phi\, L^\cc
\end{align}
is also allowed (where $L^\cc$ is an $\mathsf{SU(2)_R}$ doublet
consisting of $\left(\nu_s,e^\cc\right)$), and generates both
charged-lepton and neutrino Dirac masses.  Thus there is no symmetry
protecting sterile neutrinos from acquiring weak-scale masses in this
model.

Let us consider the other possibility, that $\mathsf{SU(2)}^\prime$ is
orthogonal to the SM.  The gauge group is
\begin{align} \label{eq:group}
  \mathsf{SU(3)_C \times SU(2)_L \times U(1)_Y \times SU(2)}^\prime \ ,
\end{align}
and the SM particles are $\mathsf{SU(2)}^\prime$ singlets.  Sterile
neutrinos lie in some representation of $\mathsf{SU(2)}^\prime$, all
of which are (almost) anomaly-free.\footnote{$\mathsf{SU(2)}$
  representations are free of gauge anomalies but may suffer from a
  global (gauge) anomaly \cite{Witten:1982fp}.  We take this into
  account shortly.}  Any half-integer representation, $\chi$, of
$\mathsf{SU(2)}^\prime$ forbids a bare mass.

Consider a single half-integer $\mathsf{SU(2)}^\prime$ representation.
The doublet representation has a global (gauge) anomaly, so it is not
a candidate \cite{Witten:1982fp}.  An even number of such
representations does not have a global (gauge) anomaly, but allows a
mass term $M^{ij}\chi_i\chi_j\ (i\not= j)$.  Thus the sterile fermions
would naturally lie at the high mass scale.

The simplest representation that is free of all anomalies is the
spin-$\frac32$ representation, which yields four sterile
neutrinos.\footnote{More generally, one may consider a model with an
  odd number of spin-3/2 representations.  Although a mass term
  $M^{ij}\chi_i\chi_j\ (i\not= j)$ is allowed, an antisymmetric mass
  matrix always has one zero eigenvalue if it is odd dimensional.
  Thus there would be one light spin-3/2 representation.}
In addition, one needs a sterile Higgs field, $\varphi$, also in the
spin-$\frac32$ representation of $\mathsf{SU(2)}^\prime$.  The
effective Lagrangian is
\begin{multline} \label{eq:l-32}
  \mathscr{L} = \frac{c_1}{M} \left(L\phi\right) \left(L\phi\right) +
  \frac{c_2}{M} \left(L\phi\right)\left(\chi\varphi\right) +
  \frac{c_3}{M} \left(L\phi\right)\left(\chi\tilde\varphi\right) +
  \frac{c_4}{M} \left(\chi\varphi\right) \left(\chi\varphi\right) +
  \frac{c_5}{M} \left(\chi\varphi\right)
  \left(\chi\tilde\varphi\right) + \frac{c_6}{M}
  \left(\chi\tilde\varphi\right) \left(\chi\tilde\varphi\right)
  \\
  + \frac{c_7}{M}\left(\chi\chi\varphi\varphi\right)
  +\frac{c_8}{M}\left(\chi\chi\varphi\tilde\varphi\right)
  +\frac{c_9}{M}\left(\chi\chi\tilde\varphi\tilde\varphi\right) +
  \text{h.c.} \ ,
\end{multline}
where $\tilde\varphi^{ijk}\equiv
\epsilon^{il}\epsilon^{jm}\epsilon^{kn}\varphi^*_{lmn}$.\footnote{The
  $\mathsf{SU(2)}^\prime$ invariants that appear are of the form
  $(\chi\phi)=\chi^{ijk}\phi_{ijk}$ and
  \mbox{$\left(\chi\chi\varphi\varphi\right) =
    \chi^{ijk}\chi^{lmn}\varphi_{ijl}\varphi_{kmn}$}.}  This model has
been analyzed in Ref.~\cite{Babu:2003is}.  It yields two light sterile
neutrinos that mix with the active neutrinos, as desired.  There are
also two other light sterile neutrinos that do not mix with the active
neutrinos, due to an unbroken $\mathbbm{Z}_3$ symmetry.\footnote{This
  is an example of a discrete gauge symmetry.  However, in this model
  it is not being used to produce a light sterile neutrino, in
  contrast to our discussion in Section~\ref{s:discrete}.}

From here, the model-building possibilities are endless.  Any model of
the form $\mathsf{G_\sm \times G^\prime}$, where $\mathsf{G}^\prime$
is a chiral, anomaly-free gauge group, can potentially yield light
sterile neutrinos along the lines of the
$\mathsf{G}^\prime=\mathsf{U(1)_X}$ and
$\mathsf{G^\prime=SU(2)}^\prime$ models discussed above. A variety of
models of this type have been analyzed in Ref.~\cite{Babu:2004mj}.
Perhaps the most compelling of these models are the mirror models,
$\mathsf{G^\prime=SU(3)_C^\prime\times SU(2)_L^\prime\times
  U(1)_Y}^\prime$ \cite{Foot:1995pa,Berezhiani:1995yi}.  While the
original versions of these models have been excluded, variations may
still be viable \cite{Berezinsky:2002fa,Foot:2002tf}. These models
yield three light sterile neutrinos.  As mentioned in the
Introduction, big bang nucleosynthesis constrains the number of light
neutrinos.  Although it may be possible to circumvent this constraint,
there is a preference for models with only a few additional neutrinos
beyond the three active ones.\footnote{For a recent review of
  neutrinos in cosmology, see Ref.~\cite{Hannestad:2004nb}.}

\section{Conclusions}

We have taken a bottom-up approach to models with naturally light
sterile neutrinos.  We have argued that there must be a symmetry
responsible for the lightness of such particles.  We considered two
classes of symmetry.  One is an exact global symmetry, broken at the
weak scale.  We found that such models are easy to construct. However,
there are serious doubts that exact global symmetries exist in nature,
due to quantum-gravitational effects.  The only model we found that
evades this criticism is based on a discrete $\mathbbm{Z}_\mathsf{N}$
symmetry.  If this symmetry is a discrete gauge symmetry, than it is
not violated by quantum gravity.  The simplest example we found is a
$\mathbbm{Z}_4$ model with two sterile neutrinos of unit charge. The
$\mathbbm{Z}_4$ symmetry is broken at the weak scale by a sterile
Higgs field of charge $3$.  We also discussed a $\mathbbm{Z}_6$ model
with sterile neutrinos of charge $(1,2)$, and a sterile Higgs field of
charge $5$.

The other class of models is based on chiral gauge theories that gives
rise to accidental global symmetries, analogous to $\mathsf{U(1)_L}$
in the standard model.  All the models we considered are of the form
$\mathsf{G_\sm \times G}^\prime$.  For $\mathsf{G}^\prime =
\mathsf{U(1)}^\prime$, the simplest anomaly-free model has five
sterile fermions, all of which are candidates for light sterile
neutrinos, depending on how the $\mathsf{U(1)}^\prime$ gauge symmetry
is broken.  We gave an example where only one light sterile neutrino
is generated that mixes with the active neutrinos. We also discussed
the case $\mathsf{G}^\prime = \mathsf{SU(2)}^\prime$, which has been
analyzed in Ref.~\cite{Babu:2004mj}.

In summary, the most attractive models of light sterile neutrinos are
based on extensions of the standard-model gauge symmetry, either by a
discrete or a continuous gauge symmetry. Thus the confirmation of the
existence of light sterile neutrinos would be evidence of a sterile
sector of particles and forces.

\begin{acknowledgements}
  We are grateful for conversations and correspondence with K.~Babu,
  J.~Conrad, B.~Fields, B.~Kayser, and H.~Murayama. This work was
  supported in part by the U.~S.~Department of Energy under contract
  No.~DE-FG02-91ER40677.
\end{acknowledgements}

\begin{appendix}

\section{Discrete Gauge Symmetry and $\mathsf{U(1)_X}$}

Although we followed a bottom-up approach throughout this paper, it is
an interesting question whether the $\mathbbm{Z}_\mathsf{N}$ models
studied in Section~\ref{s:discrete} can be obtained from a broken
gauge symmetry (discrete gauge symmetry).  Rather than addressing this
question in general, we ask the more specific question of what
$\mathbbm{Z}_\mathsf{N}$ models arise if we break the
$\mathsf{U(1)_X}$ gauge models of Section~\ref{s:u1} at the high scale
via a Higgs field of charge \textsf{N}.

As discussed in Section~\ref{s:discrete}, the simplest candidate for a
discrete gauge symmetry is $\mathbbm{Z}_4$ with two fermions of charge
unity.  However, the $\mathsf{U(1)_X}$ models listed in
Table~\ref{tb:u1-charges} lead either to four particles with unit
$\mathbbm{Z}_4$ charge, or to a vectorlike representation of
$\mathbbm{Z}_4$. Thus the $\mathbbm{Z}_4$ model discussed in
Section~\ref{s:discrete} cannot arise from any of the
$\mathsf{U(1)_X}$ models listed in Table~\ref{tb:u1-charges} (that is,
models with five commensurate charges). This does not imply, however,
that it cannot arise from some other broken gauge theory.

The other simple model we discussed in Section~\ref{s:discrete} is
$\mathbbm{Z}_6$ with two fermions of charges $(1,2)$.  This model can
result from the $\mathsf{U(1)_X}$ models listed in
Table~\ref{tb:u1-charges}.  Let's consider the simplest
$\mathsf{U(1)_X}$ model with charges $(8,7,-1,-5,-9)$ (the conjugate
of the first model listed in Table~\ref{tb:u1-charges}). If we break
the $\mathsf{U(1)_X}$ symmetry at the high scale via a Higgs field of
charge $6$, then $\psi_{-1}$ and a linear combination of $\psi_{7}$
and $\psi_{-5}$ pair up to form a massive Dirac fermion and
$\psi_{-9}$ acquires a heavy Majorana mass. The orthogonal combination
of $\psi_{7}$ and $\psi_{-5}$ as well as $\psi_{8}$ remain massless,
and these two particles indeed possess $\mathbbm{Z}_6$ charges
$(1,2)$, respectively.

\end{appendix}


\begin{thebibliography}{99}

\bibitem{Strumia:2005tc}
  A.~Strumia and F.~Vissani,
  arXiv:hep-ph/0503246.

\bibitem{Aguilar:2001ty}
  A.~Aguilar {\it et al.}  [LSND Collaboration],
  Phys.\ Rev.\ D {\bf 64}, 112007 (2001)
  [arXiv:hep-ex/0104049].

\bibitem{Weinberg:1979sa}
  S.~Weinberg,
  Phys.\ Rev.\ Lett.\  {\bf 43}, 1566 (1979).

\bibitem{Cyburt:2004yc}
  R.~H.~Cyburt, B.~D.~Fields, K.~A.~Olive and E.~Skillman,
  arXiv:astro-ph/0408033.

\bibitem{Sorel:2003hf}
  M.~Sorel, J.~M.~Conrad and M.~Shaevitz,
  Phys.\ Rev.\ D {\bf 70}, 073004 (2004)
  [arXiv:hep-ph/0305255].

\bibitem{Witten:2000dt}
  E.~Witten,
  Nucl.\ Phys.\ Proc.\ Suppl.\  {\bf 91}, 3 (2001)
  [arXiv:hep-ph/0006332].

\bibitem{Kallosh:1995hi}
  R.~Kallosh, A.~D.~Linde, D.~A.~Linde and L.~Susskind,
  Phys.\ Rev.\ D {\bf 52}, 912 (1995)
  [arXiv:hep-th/9502069].

\bibitem{Langacker:1998ut}
  P.~Langacker,
  Phys.\ Rev.\ D {\bf 58}, 093017 (1998)
  [arXiv:hep-ph/9805281].

\bibitem{Chacko:2003dt}
  Z.~Chacko, L.~J.~Hall, T.~Okui and S.~J.~Oliver,
  Phys.\ Rev.\ D {\bf 70}, 085008 (2004)
  [arXiv:hep-ph/0312267].

\bibitem{Chacko:2004cz}
  Z.~Chacko, L.~J.~Hall, S.~J.~Oliver and M.~Perelstein,
  arXiv:hep-ph/0405067.

\bibitem{Hall:2004yg}
  L.~J.~Hall and S.~J.~Oliver,
  Nucl.\ Phys.\ Proc.\ Suppl.\  {\bf 137}, 269 (2004)
  [arXiv:hep-ph/0409276].

\bibitem{Beacom:2004yd}
  J.~F.~Beacom, N.~F.~Bell and S.~Dodelson,
  Phys.\ Rev.\ Lett.\  {\bf 93}, 121302 (2004)
  [arXiv:astro-ph/0404585].

\bibitem{Shi:1993hm}
  X.~Shi, D.~N.~Schramm and B.~D.~Fields,
  Phys.\ Rev.\ D {\bf 48}, 2563 (1993)
  [arXiv:astro-ph/9307027].

\bibitem{Dolgov:2002wy}
  A.~D.~Dolgov,
  Phys.\ Rept.\  {\bf 370}, 333 (2002)
  [arXiv:hep-ph/0202122].

\bibitem{Cirelli:2004cz}
  M.~Cirelli, G.~Marandella, A.~Strumia and F.~Vissani,
  Nucl.\ Phys.\ B {\bf 708}, 215 (2005)
  [arXiv:hep-ph/0403158].

\bibitem{Beacom:2002vi}
  J.~F.~Beacom, N.~F.~Bell, D.~Hooper, S.~Pakvasa and T.~J.~Weiler,
  Phys.\ Rev.\ Lett.\  {\bf 90}, 181301 (2003)
  [arXiv:hep-ph/0211305].

\bibitem{Krauss:1988zc}
  L.~M.~Krauss and F.~Wilczek,
  Phys.\ Rev.\ Lett.\  {\bf 62}, 1221 (1989).

\bibitem{Preskill:1991kd}
  J.~Preskill, S.~P.~Trivedi, F.~Wilczek and M.~B.~Wise,
  Nucl.\ Phys.\ B {\bf 363}, 207 (1991).

\bibitem{Ibanez:1991hv}
  L.~E.~Ibanez and G.~G.~Ross,
  Phys.\ Lett.\ B {\bf 260}, 291 (1991).

\bibitem{Ibanez:1992ji}
  L.~E.~Ibanez,
  Nucl.\ Phys.\ B {\bf 398}, 301 (1993)
  [arXiv:hep-ph/9210211].

\bibitem{Appelquist:2002mw}
  T.~Appelquist, B.~A.~Dobrescu and A.~R.~Hopper,
  Phys.\ Rev.\ D {\bf 68}, 035012 (2003)
  [arXiv:hep-ph/0212073].

\bibitem{Babu:2003is}
  K.~S.~Babu and G.~Seidl,
  Phys.\ Lett.\ B {\bf 591}, 127 (2004)
  [arXiv:hep-ph/0312285].

\bibitem{Davoudiasl:2005ks}
  H.~Davoudiasl, R.~Kitano, G.~D.~Kribs and H.~Murayama,
  arXiv:hep-ph/0502176.

\bibitem{Georgi:1974sy}
  H.~Georgi and S.~L.~Glashow,
  Phys.\ Rev.\ Lett.\  {\bf 32}, 438 (1974).

\bibitem{Pati:1974yy}
  J.~C.~Pati and A.~Salam,
  Phys.\ Rev.\ D {\bf 10}, 275 (1974).

\bibitem{Witten:1982fp}
  E.~Witten,
  Phys.\ Lett.\ B {\bf 117}, 324 (1982).

\bibitem{Babu:2004mj}
  K.~S.~Babu and G.~Seidl,
  Phys.\ Rev.\ D {\bf 70}, 113014 (2004)
  [arXiv:hep-ph/0405197].

\bibitem{Foot:1995pa}
  R.~Foot and R.~R.~Volkas,
  Phys.\ Rev.\ D {\bf 52}, 6595 (1995)
  [arXiv:hep-ph/9505359].

\bibitem{Berezhiani:1995yi}
  Z.~G.~Berezhiani and R.~N.~Mohapatra,
  Phys.\ Rev.\ D {\bf 52}, 6607 (1995)
  [arXiv:hep-ph/9505385].

\bibitem{Berezinsky:2002fa}
  V.~Berezinsky, M.~Narayan and F.~Vissani,
  Nucl.\ Phys.\ B {\bf 658}, 254 (2003)
  [arXiv:hep-ph/0210204].

\bibitem{Foot:2002tf}
  R.~Foot,
  Mod.\ Phys.\ Lett.\ A {\bf 18}, 2079 (2003)
  [arXiv:hep-ph/0210393].

\bibitem{Hannestad:2004nb}
  S.~Hannestad,
  New J.\ Phys.\  {\bf 6}, 108 (2004)
  [arXiv:hep-ph/0404239].

\end{thebibliography}
\end{document}